\RequirePackage{lineno}
\documentclass[superscriptaddress,aps,preprint,amsmath,amssymb,floatfix]{revtex4-1}

\usepackage{graphicx,morefloats}
\usepackage{subfigure} 
\usepackage{color,soul}
\usepackage{bm}
\usepackage{amsmath}
\usepackage{amssymb}
\usepackage{times}
\usepackage{hyperref}
\usepackage{bbm}

\newcommand{\be}{\begin{equation}}
\newcommand{\bea}{\begin{eqnarray}}

\newcommand{\bq}{\mathbf{q}}

\newcommand{\ee}{\end{equation}}
\newcommand{\eea}{\end{eqnarray}}

\newcommand{\tbq}{\tilde{{\bf q}}}
\newcommand{\bQ}{{\bf Q}}
\newcommand{\hloc}{h_{\mathrm{loc}}}
\newcommand{\kk}{{\bf{k}}}

\newcommand{\loc}{\mathrm{loc}}

\newcommand{\SDWr}{SDW$_{\!{\rm r}}$}

\newcommand{\udb}{\overline{\uparrow\downarrow}}
\newcommand{\dub}{\overline{\downarrow\uparrow}}

\newcommand{\ud}{{\uparrow\downarrow}}
\newcommand{\du}{{\downarrow\uparrow}}

\begin{document}

\hyphenation{va-ni-sh-ing}

\begin{center}

\thispagestyle{empty}

{\large\bf 
Unconventional Superconductivity from Fermi Surface Fluctuations in Strongly Correlated Metals}
\\[0.6cm]

Haoyu\ Hu$^{1}$, Ang\ Cai$^{1}$, Lei\ Chen$^{1}$, Lili\ Deng$^{2}$, J.\ H.\ Pixley$^{3}$, Kevin Ingersent$^{2}$, and Qimiao Si$^{1,\ast}$
\\[0.3cm]

$^1$Department of Physics and Astronomy, Rice Center for Quantum Materials, Rice University, Houston, Texas 77005, USA\\[-0.cm]

$^2$Department of Physics, University of Florida, Gainesville, Florida 32611-8440, USA\\[-0.cm]

$^3$Department of Physics and Astronomy, Center for Materials Theory, Rutgers University, Piscataway, New Jersey 08854, USA\\[-0.cm]

\end{center}

\vspace{0.3cm}

{\bf 
In quantum materials, electrons that have
strong correlations tend to localize, leading to quantum spins as the building blocks for low-energy physics \cite{Kei17.1,Pas21.1}. 
When strongly correlated electrons coexist with more weakly-correlated conduction electrons, 
multiple channels of effective interactions develop and compete with each other.
The competition creates quantum fluctuations having a large spectral weight, with the associated entropies reaching significant fractions of 
$R\ln 2$  per electron. 
Advancing a framework to understand how
the fluctuating local moments influence unconventional superconductivity
 \cite{Shr21.1,park2006hidden,stockert2011magnetically} is both pressing and challenging.
Here we do so in the exemplary setting of heavy-fermion metals,
where the amplified quantum fluctuations manifest in the form of Kondo destruction and large-to-small
Fermi-surface fluctuations. These fluctuations lead to unconventional superconductivity 
whose transition temperature is exceptionally high relative to the effective Fermi temperature,
reaching several percent of the Kondo temperature scale.
Our results provide a natural understanding of the enigmatic superconductivity in a host of heavy-fermion metals. 
Moreover, the qualitative physics underlying our findings and their implications for the formation 
of unconventional superconductivity
apply to a variety of 
highly correlated metals with strong Fermi surface fluctuations
\cite{Bad16.1,Oik15.1}.
}
\vspace{0.6cm}

\noindent E-mail: $^{\ast}$qmsi@rice.edu

\newpage
Strong correlations 
drive a 
plethora
 of quantum phases \cite{Kei17.1,Pas21.1}. 
Heavy-fermion systems represent a prototype of strongly correlated metals \cite{Col05.1,Kir20.1}.
Here, the
$f$ electrons
have a Coulomb repulsion much larger than their kinetic energy, and
at low energies they
act as quantum spins. 
The spins are coupled to the $spd$-based conduction electrons
by an antiferromagnetic (AF) 
exchange, the Kondo interaction,  and a
Ruderman-Kittel-Kasuya-Yosida (RKKY) interaction 
between the spins that is typically AF as well.
The Kondo interaction promotes a ground state with a nonzero amplitude for a collective
spin singlet between the local moments and conduction electrons.
The Kondo energy scale acts as an effective Fermi energy for
the composite fermions, the reincarnation of the local moments that are fractionalized as a result of 
their inter-locking with
the charge-carrying conduction electrons. 
Unconventional superconductivity develops in
about 50 heavy-fermion superconductors,
in many cases close to an AF-ordered phase.
Examples include
CeRhIn$_5$, which is a part of the Ce-115 materials family with $T_c \approx 2.3$ K
(a record high among $4f$-electron-based heavy-fermion systems),
and CeCu$_2$Si$_2$, which has $T_c \approx 0.6$K and is the very first unconventional superconductor ever discovered.
These transition temperatures are exceptionally high, recognizing that their
ratio to the respective effective Fermi temperature is a few percent.
This is to be contrasted with what happens in conventional superconductors, 
where the ratio is typically 
orders of magnitude smaller.

There is ample empirical evidence that strong correlations, in the form of the Kondo effect, are key to the 
development of heavy-fermion superconductivity. 
The amount of entropy involved in the superconducting condensation
is a sizeable
 fraction of R$\ln$2  per electron,
implying that spin-$\frac{1}{2}$ local moments are active agents for the superconductivity.
A host of 
spectroscopic measurements support this
 perspective \cite{Shr21.1}.
With a few exceptions \cite{Flint2008,Wu15.1}, the Kondo effect has not been incorporated
into
theoretical studies of the mechanism
of unconventional heavy-fermion superconductivity.

\begin{figure}[h!]
\center
\includegraphics[height=6.3in]{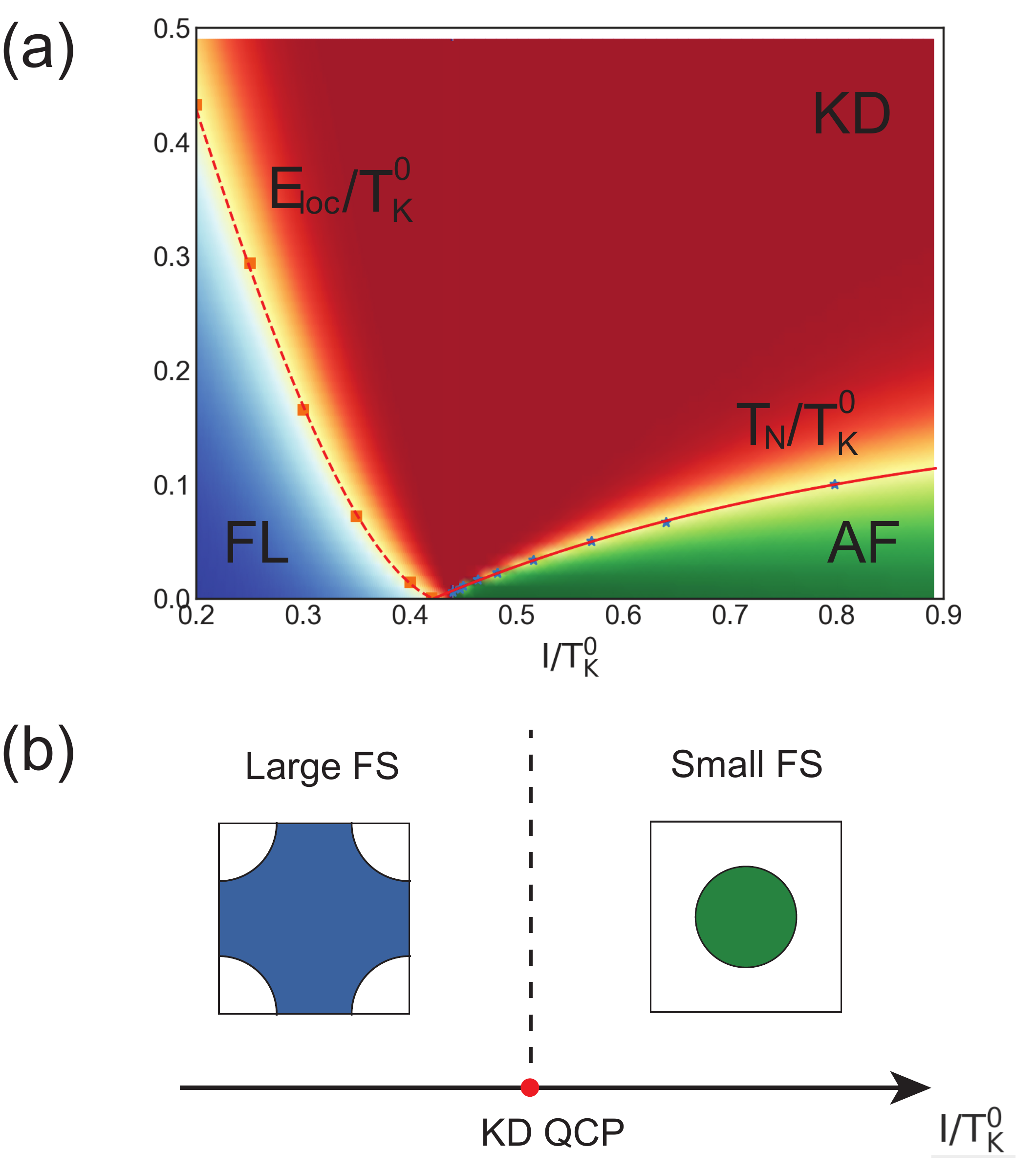}
\caption{(a) Finite-temperature phase diagram vs $I/T_K^0$ for 
type I model. For small $I/T_K^0$, the model is in 
a heavy Fermi liquid (FL) phase characterized by the static spin susceptibility
saturating to a constant for   
$T \ll E_{\loc}$.  
For large $I/T_K^0$, the model develops antiferromagnetic (AF) order.  
Between these two regions,  in a ``fan'' above the point where $E_{\loc}$
goes to zero at the QCP, is the quantum critical
Kondo destruction (KD) regime of non-Fermi liquid behavior.
(b) Schematic reconstruction of the Fermi surface from large to small through the KD QCP.
In the quantum critical regime, the Fermi surface fluctuates between the two.
}
\label{fig:phase_diagram}
\end{figure}

What has been especially lacking is a framework for how unconventional superconductivity 
develops out of heavy-fermion quantum criticality \cite{Col05.1,Kir20.1}, which arises from 
a dynamical competition between the RKKY and Kondo
 interactions and
has been recognized as a central ingredient of the correlation physics in the normal state.
The AF RKKY interactions, which boost spin-singlet correlations among the local moments, 
weaken the static Kondo-singlet amplitude.
This can lead to two types of AF QCPs \cite{si2001locally,coleman2001fermi,Sen04.1}.
In one type, the Kondo destruction (KD) QCP,
the static amplitude of the Kondo singlet vanishes as the QCP is approached from the paramagnetic side.
The
$f$ electrons go from being itinerant composite fermions,
which participate in the Fermi-surface formation, to being localized in the AF ground state.
This large-to-small Fermi-surface reconstruction characterizes a
partial Mott (delocalization-localization) transition
across the QCP (see Fig.~\ref{fig:phase_diagram}), 
and is a key part of the experimental 
signatures \cite{Pro20.1,paschen2004hall,friedemann2010fermi,shishido2005drastic,park2006hidden}.
In the other
type of quantum criticality, the Kondo amplitude remains nonzero at the QCP
and the heavy quasiparticles undergo a spin density wave (SDW) transition.
Given that the immediately adjacent AF order is an SDW formed from the 
{\it renormalized} $f$-electron-based composite quasiparticles, we will refer to this
as an \SDWr\ QCP to distinguish it from a conventional SDW transition.

Here we address unconventional superconductivity arising from
KD and the concomitant 
delocalization-localization transition of the
$f$ electrons.
The quantum fluctuations, which have a large spectral weight,
are found to 
yield robust spin-singlet superconductivity, 
with $T_c$ reaching a few percent of the 
Kondo temperature scale.
Although our analysis is focused on a concrete model suitable for heavy-fermion
systems,
quantum criticality associated with a delocalization-localization transition
appears broadly relevant to a variety of other strongly correlated metals \cite{Pas21.1}.
Possible materials classes in this category include high-temperature cuprate
superconductors \cite{Bad16.1}, organic charge-transfer salts \cite{Oik15.1},
and moir\'{e} systems \cite{Ghiotto2021,Li2021}.

A canonical microscopic model for heavy-fermion systems is the Anderson lattice model. It
describes a single band of conduction electrons hybridizing with strength $V$
with a band of
$4f$ electrons
that have a
strong on-site Hubbard interaction $U$.  
The large $U$ creates an antiferromagnetic Kondo
exchange coupling $J_K \propto V^2/U$ between
local $f$ moments and itinerant conduction electrons. Acting alone, this
gives rise to
a bare Kondo temperature 
$T_K^{0}$ [$\approx \rho_0^{-1} \exp{-(1/\rho_0 J_K)}$, with $\rho_0$ 
being the bare conduction-electron density of states at the Fermi energy],
below which the local moments
are screened by the conduction electrons.
The RKKY interaction $I$ acts between the localized magnetic moments.  
The ratio
$\delta \equiv I/T_K^{0}$ determines whether the system will order magnetically or develop heavy Fermi liquid behavior.
The periodic Anderson
Hamiltonian is (see Methods)
\begin{eqnarray}
H_{\mathrm{AL}} &=& \sum_{
 i,j, \sigma} t_{ij}(c_{i\sigma}^{\dag}c_{j\sigma} + \mathrm{h.c}) + \sum_{i} \left(\epsilon_f n_{fi} +U n_{fi\uparrow}n_{fi\downarrow}\right)
\nonumber
\\
&+&
 \sum_{i,\sigma}\left(Vc_{i\sigma}^{\dag}f_{i\sigma} + \mathrm{h.c.} \right) + \sum_{
  i,j,m} I^{m}_{ij}S^{m}_{fi} S^{m}_{fj} \, ,
\label{eqn:ham}
\end{eqnarray}
where
$c_{i\sigma}$ ($f_{i\sigma}$) 
destroys a conduction ($4f$) electron at lattice site $i$ with spin $\sigma$,
while $\epsilon_f$ and $U$ are respectively the $f$-level energy and on-site Coulomb repulsion.
In addition to a $c$-electron hopping $t_{ij}$ between
lattice sites
$i,j$, 
we have explicitly included an
RKKY exchange
$I^{m}_{ij}$ between Cartesian component $m \in \{x, y, z\}$ of the localized
$4f$ moments.
We focus on RKKY interactions in the limits of either Ising anistropy 
($I^x_{ij}=I^y_{ij}=0$, $I^z_{ij}=I_{ij}$)
or full SU(2) symmetry ($I^m_{ij}=I_{ij}$).
It is also important to distinguish two types of model according to the way in which the
``RKKY density of states'' $\rho_I(\epsilon) \equiv \sum_{\bq} \delta(\epsilon - I_{\bq})$ increases
from its lower edge at $\epsilon=I_{\bQ}$. (Here, $I_{\bq}$ is the Fourier transform of $I_{ij}$ and
$\bQ$ is the ordering wave vector in the magnetic
phase.)
In type I models, 
$\rho_I(\epsilon)$ has a jump onset, characteristic of two-dimensional magnetic fluctuations.
In type II models, $\rho_I(\epsilon)$ instead increases
smoothly $\propto \sqrt{\epsilon-I_{\bQ}}$,
reflective of three-dimensional magnetic fluctuations.
The Anderson lattice model has been studied in a variety of contexts. For quantum phase transitions, the
 distinction between KD and \SDWr\ criticality has been explored
through an extended dynamical mean-field theory (EDMFT) approach \cite{si2001locally,si2003local},
with the most detailed results obtained for the case of Ising symmetry
\cite{grempel2003locally,zhu2003continuous,glossop2007magnetic,zhu2007zero}.
The quantum critical dynamics of the KD QCP plays an important role in connecting the theory to
experiments \cite{schroder2000onset}.

In order to permit study of unconventional superconducting pairing, which is necessarily
off-site, we here report the first application of a cluster EDMFT (C-EDMFT) \cite{pixley2015cluster}, 
which maps $H_{AL}$ to a self-consistently determined two-site quantum cluster model 
(see Methods and ref.\ \citenum{pixley2015cluster} for additional details).
We solve the effective cluster model using
the numerically exact 
continuous time quantum Monte Carlo (CT-QMC) method \cite{gull2011continuous} 
at nonzero temperatures 
in a form suitable for our purpose
\cite{pixley2013quantum,pixley2015pairing,cai2019bose}.

\begin{figure}[h!]
\center
\includegraphics[width=5.8in]{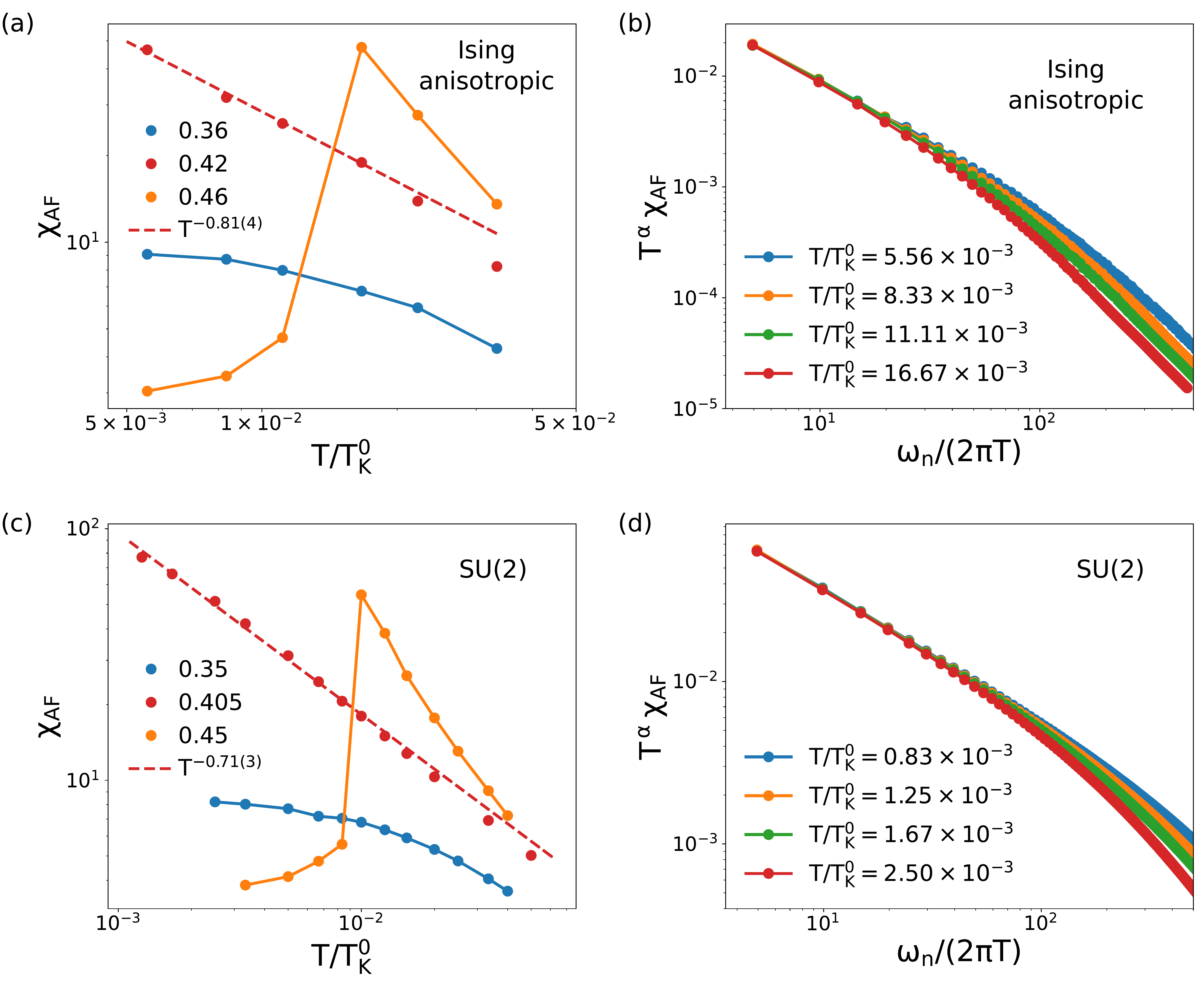}
\caption{(a) Temperature dependence of $\chi_{AF}$, the static lattice spin susceptibility 
at the ordering wave vector, and (b) demonstration of $\omega_n/T$ scaling for the dynamical
lattice spin susceptibility, at the KD QCP in the Ising-anisotropic model. Their counterparts
for the SU(2) case are shown in (c) and (d), respectively.
}
\label{fig:static-and-dyn-chi_AF}
\end{figure}

To assess the
ability of the C-EDMFT approach to properly capture the quantum critical dynamics, we
first consider the type I model in the limit of Ising-anistropic RKKY interactions.
We identify a KD QCP, as demonstrated by the phase diagram
in Fig.\,\ref{fig:phase_diagram}(a). 
Starting from the paramagnetic side,
as the tuning parameter $\delta \equiv I/T_K^0$ is increased, a renormalized Kondo scale $E_{\loc}$ 
vanishes at the continuous onset of 
AF order. The suppression of this energy scale implies the destruction of the Kondo
resonance---often referred to
as composite fermions or simply
$f$ fermions---thereby leading to a transformation from a large Fermi surface (incorporating the 
$f$ fermons) to a small one (excluding the
$f$ fermions), as illustrated schematically in Fig.\ \ref{fig:phase_diagram}(b).
At the KD QCP, the temperature dependence of the AF spin susceptibility has a power-law dependence, 
\begin{equation}
\chi_{AF}(T)\sim T^{-\alpha}
\label{eqn:chiaf}
\end{equation}
with a fractional exponent
$\alpha_{\mathrm{Ising}}=0.81(4)$ [Fig.\ \ref{fig:static-and-dyn-chi_AF}(a)], and obeys $\omega_n/T$
scaling
[Fig.\ \ref{fig:static-and-dyn-chi_AF}(b)]. 
These
are essentially the same
results as obtained for Ising anisotropy via single-site EDMFT
\cite{grempel2003locally,zhu2003continuous,glossop2007magnetic,zhu2007zero},
which captures the fractional exponent $\alpha$
in the dynamical spin susceptibility that has been measured by inelastic neutron scattering
\cite{schroder2000onset}
in
Ising-anisotropic
CeCu$_{5.9}$Au$_{0.1}$.
The C-EDMFT calculation demonstrates the robustness of the KD QCP in the presence of finite-size
corrections,
an important finding
given that the anomalous dynamical scaling of the spin response is a key signature of this type of QCP.

\begin{figure}[h!]
\center
\includegraphics[height=5.0in,angle=0]{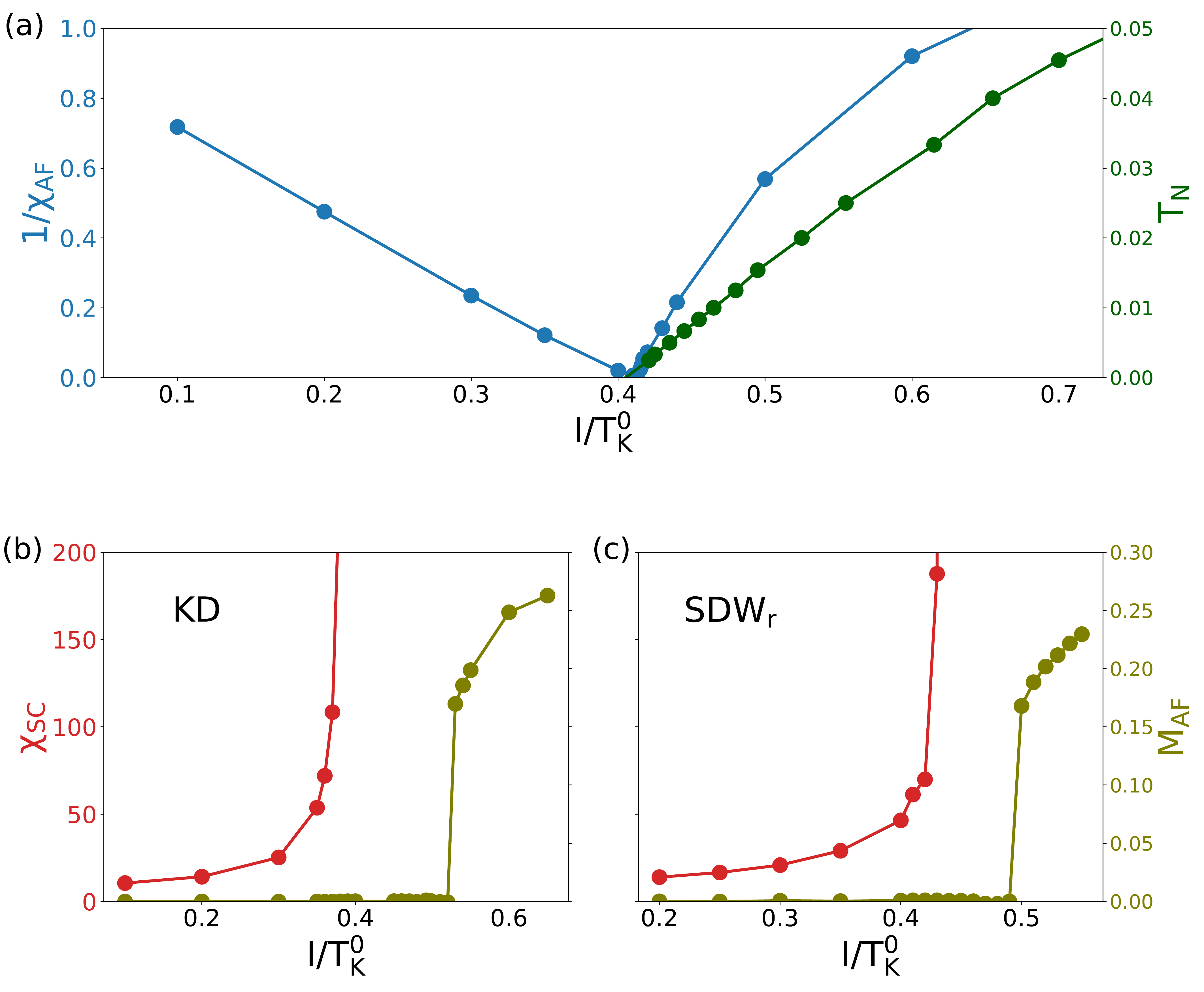}
\caption{(a) Variation with $I/T_{K}^{0}$ of the inverse static lattice spin susceptibility at the ordering
wave vector, $1/\chi_{AF}$, calculated at the lowest accessed temperature $T/T_{K}^{0}=0.001$,
and of the AF transition temperature $T_N$.
(b), (c) Static lattice pairing susceptibility $\chi_{SC}$ and
AF order parameter ${M_{AF}}$ vs $I/T_{K}^{0}$ in the type I and type II models, respectively,
at a relatively high temperature $T/T_{K}^{0} = 0.02$.
}
\label{fig:chi_SC_and_M}
\end{figure}

In the context of
unconventional pairing and superconductivity,
we expect
the spin-flip part of the RKKY interaction 
to be essential
for driving the formation of spin-singlet
Cooper pairs~\cite{pixley2015pairing,cai2016critical,Hu21.1x}.
Accordingly, we
have solved the C-EDMFT equations of the type I model with SU(2) symmetry.
Fig.\ \ref{fig:chi_SC_and_M}(a) shows the inverse of the static lattice spin susceptibility $\chi_{AF}$ obtained 
at the lowest temperature $T/T_{K}^{0}=0.001$
and the AF transition temperature 
$T_N$, both as functions
of the tuning parameter $I/T_{K}^{0}$.
It is seen that $T_{N}$
goes continuously to zero at the QCP $I/T_K^0=I_{c}/T_K^0\simeq 0.405(5)$,
where $\chi_{AF}$ also diverges. 
This provides evidence that in the SU(2) symmetric case,
just as for Ising anisotropy, the zero-temperature transition is second
order.
The quantum-critical behavior is also of the KD type: At the QCP, $\chi_{AF}$ follows Eq.\ \eqref{eqn:chiaf} with 
a fractional exponent 
$\alpha_{SU(2)}=0.71(3)$
[Fig.\ \ref{fig:static-and-dyn-chi_AF}(c)],
and it obeys an $\omega_n/T$ dynamical scaling with the same fractional exponent
[Fig.\ \ref{fig:static-and-dyn-chi_AF}(d)]. 
These properties are similar to the single-site EDMFT solution \cite{Hu20.1x}, including the value of the
exponent $\alpha$.
Our C-EDMFT results demonstrate the robustness of the Kondo-destruction nature of the quantum-critical properties
in the normal state of both the Ising and SU(2) limits of the type I model.
 
We are now in position to study 
pairing correlations. 
The pairing susceptibility $\chi_{SC}$ in the spin-singlet channel
[see Supplementary Information, Eq.\,(\ref{eq:chisc_real})]
diverges, demonstrating a superconducting phase below
a transition temperature $T_{c}$. 
This is illustrated in Fig.\ \ref{fig:chi_SC_and_M}(b),(c) 
which plot the $I/T_K^0$ dependence of
$\chi_{SC}$ and the AF order parameter at $T/T_{K}^{0}=0.02$.
It is seen that $\chi_{SC}$ diverges 
as the AF transition is approached from the paramagnetic
side, becoming infinite at some $I/T_K^0 < I_c/T_K^0$.

\begin{figure}[h!]
\center
\includegraphics[height=2.6in,angle=0]{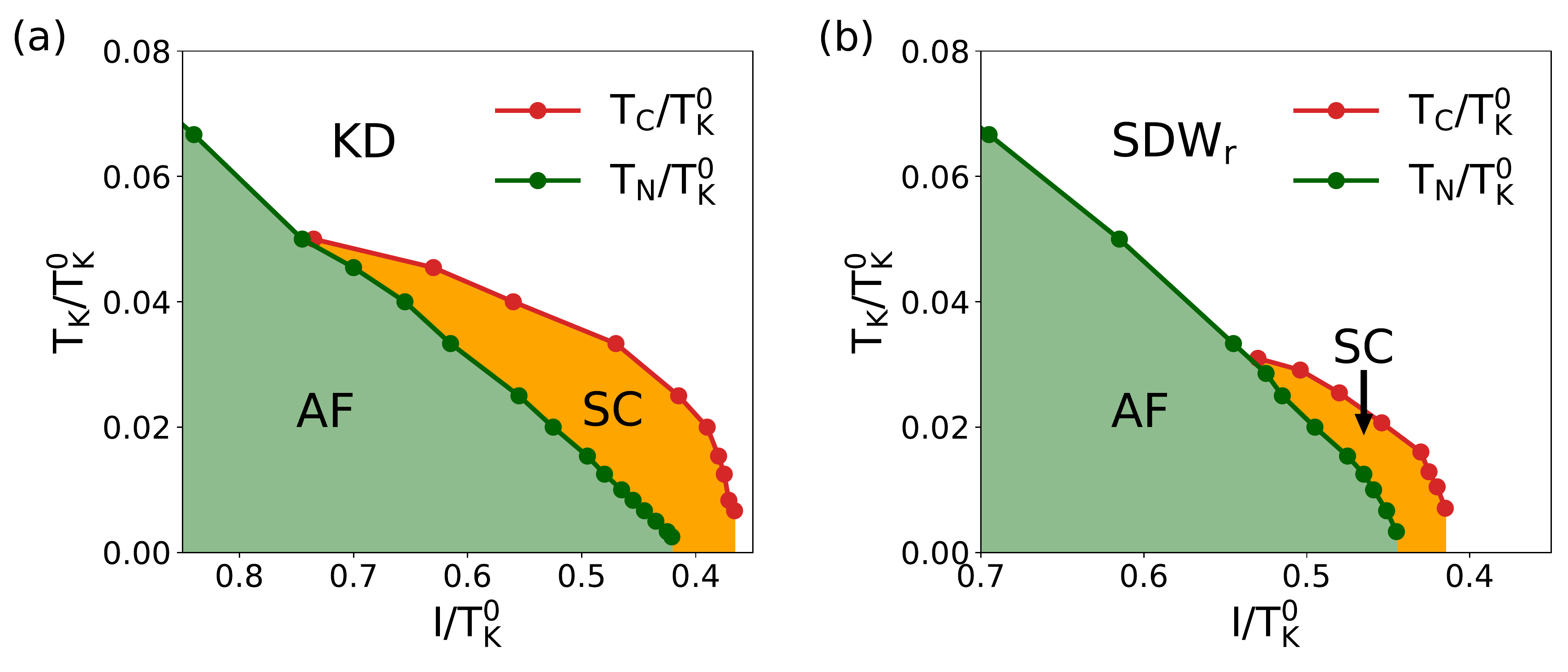}
\caption{Finite-temperature phase diagram vs control parameter $I/T_{K}^{0}$ for (a) KD and (b) \SDWr\ quantum criticality
in the SU(2) symmetric case. 
The transition temperature for the antiferromagnetic (AF) and superconducting (SC) phases is marked by green and red circles,
respectively.
}
\label{fig:T_SC_SU2}
\end{figure}

We have carried out such calculations at various temperatures 
and
used the location where $\chi_{SC}$ diverges to determine the finite-temperature phase boundary 
for the superconducting phase. The obtained phase
in Fig.\ \ref{fig:T_SC_SU2}(a) shows
a broad region of superconducting order 
near to, and indeed hiding, the QCP.
The superconducting transition temperature $T_{c}$ reaches a maximum 
of about 5\% of the bare Kondo temperature $T_{K}^{0}$; its value at the QCP is about
$0.03 T_K^0$.

We next turn to the second type of quantum critical solution, which we derive in the type II model
and is of the \SDWr\ type (see
Supplementary Information).
Here the
renormalized Kondo energy scale $E_{\loc}$ does not go to zero upon reaching 
the QCP from the paramagnetic side.
However, $E_{cr}$---the value of 
$E_{loc}$
at the QCP---is small compared to the bare 
Kondo energy scale $k_BT_K^0$
(Supplementary Information, Fig.\,\ref{fig:Ising-sdw}).
The asymptotic behavior at energies below $E_{cr}$ takes the form of the quantum criticality associated with the 
conventional
SDW QCP \cite{hertz1976quantum,millis1993effect}.
The small $E_{cr}/k_B T_K^0$
reflects the considerable reduction of
Kondo-singlet correlations
due to AF correlations between local moments, and serves to distinguish \SDWr\ quantum criticality
from conventional SDW QCPs \cite{hertz1976quantum,millis1993effect}
where the Kondo effect does not 
operate.

At
an \SDWr\ QCP,
quantum fluctuations
in the intermediate energy range $E_{cr} \lesssim E \lesssim k_BT_K^0$
still manifest the physics of
disintegrating Kondo singlets, i.e., the delocalization-localization transition with the Fermi surface
crossing over from large to small.
These fluctuations 
will drive unconventional superconductivity
along the same lines as at a KD QCP,
albeit
with a more limited dynamical range and a weaker pairing strength. Indeed, we find that spin-singlet 
superconductivity
develops in the type II model [Fig.\,\ref{fig:T_SC_SU2}(b)]
with a lower $T_c$ than in the type I model
($\simeq 0.02 T_K^0$ at the QCP compared with
$0.03 T_K^0$).

Our work provides a very general understanding of the superconducting pairing in quantum-critical heavy-fermion metals. 
We illustrate this point in the context of 
two prominent examples.
The first is CeRhIn$_5$, which
features 
a pressure-induced QCP \cite{park2006hidden,Kne08.1,Kir20.1,Tho12.1}:
As illustrated in Fig.\,\ref{fig:exp}, superconductivity develops near the QCP where 
$T_c \approx 2.3$K, which is about 5\% of the bare Kondo temperature ($\approx 40$ K; ref.\ \citenum{Kir20.1}). 
In the magnitude of $T_c$
as well as in
the spin-singlet nature of the pairing and other properties of its superconducting state, 
CeRhIn$_5$ is similar to its high-chemical-pressure counterpart CeCoIn$_5$ under ambient conditions \cite{Shr21.1}.
In both cases, the spin anisotropy is relatively small \cite{Kir20.1},
i.e., the SU(2) limit should apply. Importantly, across the critical pressure 
of CeRhIn$_5$, quantum oscillation measurements have established a small-to-large Fermi surface jump \cite{Shi05.1,Kir20.1},
providing strong evidence for
the KD nature of the QCP. Our finding of
superconductivity at the KD QCP with a high $T_c/T_K^0$ 
provides the first theoretical understanding of how superconductivity develops from a strange metal in CeRhIn$_5$.
 
A second
important material is CeCu$_2$Si$_2$, the very first unconventional superconductor
observed in nature \cite{Smidman}.
Here, there is considerable experimental evidence that the quantum criticality is of the \SDWr\ type,
with a crossover temperature $T_{cr}=E_{cr}/k_B \approx 1$ K
that is small compared to the bare Kondo
temperature ($T_K^0 \approx 20$ K).
The spin damping rate crosses over from $\propto T^{3/2}$ for $T<T_{cr}$ to $\propto T$ for $T>T_{cr}$
(Ref.\ \citenum{Smidman,Arndt11}).
Further evidence for 
the involvement of a $T_{cr}$ that is small compared to $T_K^0$ has come from an estimate of a change in the kinetic energy across the superconducting 
transition \cite{stockert2011magnetically}. We therefore advance the notion that superconductivity in CeCu$_2$Si$_2$ is driven by \SDWr\ 
quantum criticality 
that sets in below a decade of energies dominated by KD quantum fluctuations.
 
In our approach, the dynamical competition between the RKKY and Kondo couplings plays a crucial role in 
capturing the dynamical $\omega/T$ scaling and fractional exponent as well as the small-to-larger Fermi surface transformation
in the quantum critical regime,
from which
 the superconducting state
 develops.
This is to be contrasted with other theoretical approaches.
One scenario considers multiple channels of conduction ($c$) electrons
with the $f$-electron-derived local moments. The multi-channel Kondo effect is suggested to yield a composite
$f$-$c$ pairing \cite{Flint2008}. Because
RKKY interactions are not involved, this picture does not associate the superconductivity with
a quantum-critical normal state. An alternative
scenario invokes a cluster of local moments self-consistently Kondo-coupled to 
a conduction-electron bath \cite{Wu15.1}. In this picture, the AF order comes from a static (Hartree-Fock) treatment
of RKKY interactions; the lack of dynamical RKKY-Kondo competition also implies the absence of the
 partial-Mott ($f$-electron delocalization-localization) effect 
  in the normal state.

Taking a wider perspective,
we have already noted that a delocalization-localization transition 
and the accompanying small-to-large Fermi surface transformation
have been implicated in a broad range of strongly correlated metals \cite{Pas21.1}.
These in particular include doped Mott insulators \cite{Bad16.1,Oik15.1},
where a rapid change in the carrier concentration has been observed near optimal superconductivity,
Analysis of dynamical equations 
related to the single-site EDMFT analysis of the Kondo lattice has recently been carried out for doped Mott insulators
\cite{Cho21.1x}. It clearly is important to address whether unconventional superconductivity develops in
the dynamical equations for
such a setting.

To summarize, we have developed a framework to address superconductivity that develops out of a quantum-critical normal state
featuring a partial Mott (delocalization-localization) transition and small-to-large Fermi surface transformation.
Our theoretical approach captures the critical dynamics in a robust way, and demonstrates superconductivity whose transition
temperature is as high as a few percent of the effective Fermi energy.
The results provide the first natural understanding of unconventional superconductivity in a variety of prominent
families of heavy-fermion materials and
offer a promising
framework for interpreting superconductivity in
other strongly correlated metals such as doped Mott insulators.

\bibliographystyle{naturemagallauthors}
\bibliography{CEDMFT_reference}

\clearpage

\vspace{0.3cm}
\noindent{\bf Acknowledgments}\\
We would like to thank Gabriel Aeppli, Piers Coleman, Laura H. Greene, Kazushi Kanoda, Stefan Kirchner, Gabriel Kotliar, Silke Paschen, 
Frank Steglich, Oliver Stockert, Joe D. Thompson, Huiqiu Yuan and Jian-Xin Zhu 
for useful discussions. This work was supported in part by 
the National Science Foundation under Grant No.\ DMR-1920740 (H.H., L.C., and Q.S.)
and the Robert A. Welch Foundation Grant No.\ C-1411 (A.C.). 
Work at Rutgers University was supported by the Alfred P.
Sloan Foundation through a Sloan Research Fellowship (J.H.P.).
Computing time was allocated in part by the Data Analysis and Visualization Cyberinfrastructure 
funded by NSF under grant OCI-0959097 and an IBM Shared University Research (SUR) Award at Rice University, 
and by the Extreme Science and Engineering Discovery Environment (XSEDE) by NSF under Grants No. DMR170109.
J.H.P. acknowledges the hospitality of Rice University. J.H.P., K.I., and Q.S. acknowledge the hospitality of the Aspen Center for Physics,
which is supported by the NSF under Grant No.\ PHY-1607611.

\vspace{0.2cm}
\noindent{\bf Author contributions}
The first two authors contributed equally to this work. 
All authors contributed to the research of the work and the writing of the paper. 

\vspace{0.2cm}
\noindent{\bf Competing 
 interests}\\
The authors declare no competing 
 interests.
 
 \vspace{0.2cm}
 \noindent{\bf Additional information}\\
Correspondence and requests for materials should be addressed to 
Q.S. (qmsi@rice.edu)

\clearpage

\clearpage

\clearpage

\clearpage

\clearpage

\noindent{\bf\large Methods}\\
\\
\noindent
{\bf C-EDMFT method}~~
We solve the periodic Anderson model [Eq.\,(\ref{eqn:ham})] using the C-EDMFT approach of Ref.\,\onlinecite{pixley2015cluster}, focusing
on a two-site cluster that allows us to study the formation of unconventional Cooper pairs.
  We iteratively solve the C-EDMFT equations seeking self consistency. 
 In order to 
 achieve reasonably high accuracy, we solve the cluster model at finite temperature using the numerically exact CT-QMC method.
Away from the critical regime, 
the self-consistency loop converges quite fast, 
within about $30$ iterations.
Near the critical point, however, there is a critical slowing down.
We find it useful to employ simple mixing techniques~\cite{Zitko-2009} to accelerate convergence.
Still, the number of iterations can become very large (even exceeding $1000$).  

\noindent
{\bf Quantum phase transition}~~
We concentrate on the static spin susceptibility $\chi_{AF}(T)$ at the 
AF ordering wave vector $\bQ_{\mathrm{AF}}$
(e.g.
$\bQ_{\mathrm{AF}}=(\pi/a,\pi/a)$ for a lattice spacing equal to $a$ in two dimensions)
and temperature $T$.  
We mark entry into the antiferromagnetically ordered phase  through a diverging $\chi_{AF}$ and 
a nonzero order parameter $M_{AF}$.
To determine the fate of the
renormalized Kondo 
energy scale $E_{\loc}$,
 we also consider the local spin susceptibility in the 
 AF
 channel $\chi_{\loc}(\bQ_{\mathrm{AF}},i\omega_n,T)$, where $\omega_n = 2\pi n  k_B T/\hbar$ 
 denotes the Matsubara frequency.  For zero RKKY interaction, a measure of the effective single ion Kondo temperature 
 $T_K^0$ can be determined through the inverse of $\chi_{\loc}(\bQ_{\mathrm{AF}},i\omega_n=0,T=0)$
  (see Supplementary Information).
 The renormalized Kondo energy scale $E_{\loc}$ is related to the dynamical local spin susceptibility
 (see Supplementary Information and Ref.\,\onlinecite{si2001locally,si2003local}). 
 All spin susceptibilities are measured in units of $(g\mu_B)^2$, 
 where $g$ is the Land{\' e} $g$-factor of the local moment and $\mu_B$ is the Bohr magneton.

\noindent
{\bf Superconducting instability}~~
To investigate the superconducting instability, we calculate the static pairing susceptibility in the spin-singlet channel $\chi_{SC}(T)$ via 
a Bethe-Salpeter equation. The superconducting transition temperature is determined by the divergence of the pairing susceptibility 
$\chi_{SC}(T\rightarrow T_c)\rightarrow \infty$.

\noindent
{\bf Data availability}~~
The data that support the findings of this study are available from the corresponding author
upon reasonable request.

\noindent
{\bf Code availability}~~
The relevant codes used in this study are available from the corresponding author
upon reasonable request.

\clearpage
\begin{figure}[h!]
\center
\includegraphics[height=3.6in,angle=0]{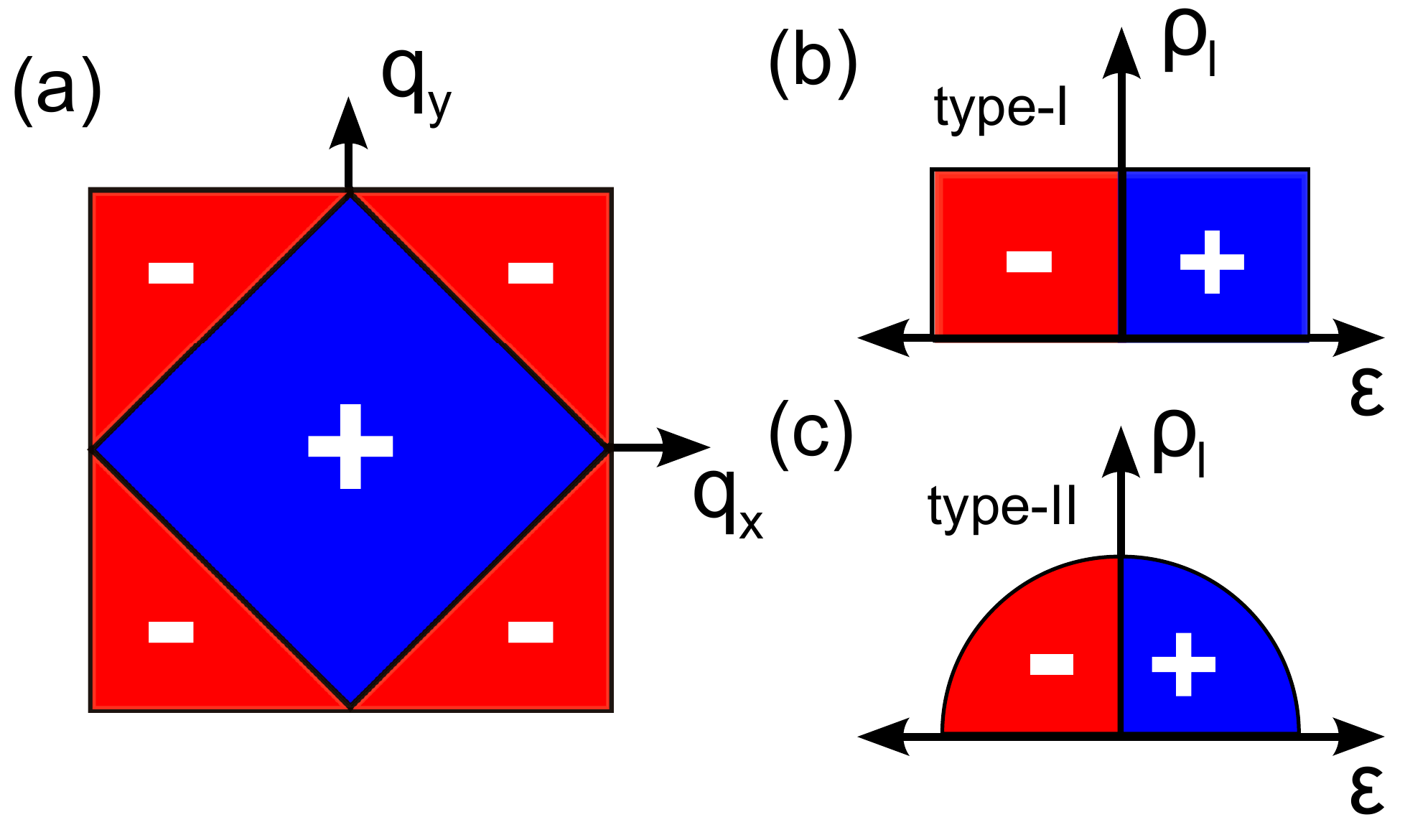}
\caption{Tiling of the
Brillioun zone with the two site cluster (a), where the ferromagnetic fluctuations are confined to the zone center [the blue region (+)] 
and the antiferromagnetic region at the zone corners [the red region (-)].  
The 
RKKY density of states in the type-I (b) and type-II (c) models are reflective of the two- and three-dimensional magnetic fluctuations.
 }
 \label{fig:rkky-dos}
\end{figure}

\clearpage

\setcounter{figure}{0}
\setcounter{equation}{0}
\makeatletter
\renewcommand{\thefigure}{S\@arabic\c@figure}
\renewcommand{\theequation}{S\arabic{equation}}

\noindent{\bf\Large Supplementary Information}\\
\\
{\bf C-EDMFT method}~~
The effective RKKY density of states is defined as follows:
\begin{equation}
\rho_I(\epsilon) = \sum_{\bq} \delta(\epsilon - I_{\bq})
\end{equation} 
where $I_{\bq}$ is the Fourier transform of $I_{ij}$. 
Using the two-site cluster, we tile the Brillioun zone so that all ferromagnetic fluctuations are confined to the zone center  
with cluster momentum $\bQ = \bQ_+$, and the antiferromagnetic fluctuations on the zone corners with cluster momentum $\bQ = \bQ_-$, 
see 
Fig.\,\ref{fig:rkky-dos}.
In the following we use $\bQ$ and $\tbq$ to denote inter and intra cluster momentum respectively
and the ordering wavevector is given by $\bQ_{AF}=\bQ_-$.
This tiling of the Brillioun zone leads to a density of states per patch, 
with $\rho_I(\epsilon) = \sum_{\tbq, \bQ } \delta(\epsilon - I_{\tbq+\bQ}) =  \sum_{\bQ}\rho_I(\bQ,\epsilon)$ 
and we have defined the density of states in patch $\bQ$.
\begin{equation}
\rho_I(\bQ,\epsilon) = \sum_{\tbq} \delta(\epsilon- I_{\tbq+\bQ}).
\end{equation}
In the Ising anisotropic case, we choose the cluster momentum in two dimensions $\bQ_+ = (0,0)$ and $\bQ_- = (\pi,\pi)$, 
as well as $\bQ_+ = (0,0,0)$ and $\bQ_- = (\pi,\pi,\pi)$ 
in three dimensions (note that the antiferromagnetic ordering wave vector corresponds to $\bQ_{AF} = \bQ_-$).  
We set the RKKY interaction $I_{ij}$ to be 
AF and only consider nearest neighbor interaction, i.e. $I_{q}=I(\cos(q_{x}) + \cos(q_{y}) )$ for 
the type I case and $I_{q}=\frac{2}{3} I(\cos(q_{x})+\cos(q_{y})+\cos(q_{z}) )$ for the type II case. 
(We always take $I_{\bQ_{AF}}=-2I$, which serves as our definition for $I$). We approximate the two dimensional density 
of states with a jump at the lower zone edge. Whereas the three dimensional magnetic density of states 
vanishes in a square root fashion at the lower zone edge (see Fig.\,\ref{fig:rkky-dos}). 
\begin{eqnarray}
\rho^{\rm type-I}_{I}(\bQ_{\pm},\omega) &=& \frac{1}{2I} \Theta (\pm \omega) \Theta (2I \mp \omega)
 \label{eq:rho2d} \\
\rho^{\rm type-II}_{I}(\bQ_{\pm},\omega) &=& \frac{1}{\pi I^{2} } \Theta(\pm \omega) \Theta ( 2I \mp \omega) \sqrt{(2I)^{2}-\omega^{2}} \label{eq:rho3d} 
\end{eqnarray}

In the SU(2) case,
we also incorporate the next nearest neighbor RKKY interaction.
For 
the type-I case we have $I_{q}=I_{1}(\cos(q_{x})+\cos(q_{y}))+I_{2}(\cos(q_{x}+q_{y})+\cos(q_{x}-q_{y}))$. 
We choose $I_{1}=-1.5I_{2}>0$ and the cluster momentum $\bQ_+ = (0,0)$ and $\bQ_- = (\pi,\pi)$. 
For
the type II case we have $I_{q}=I_{1}(\cos(q_{x})+\cos(q_{y})\cos(q_{z}))
+I_{2}(\cos(q_{x}+q_{y})+\cos(q_{x}-q_{y})+\cos(q_{y}+q_{z})+\cos(q_{y}-q_{z})+\cos(q_{z}+q_{x})+\cos(q_{z}-q_{x}))$. 
We choose $I_{1}=0.65I_{2}>0$ and the cluster momentum $\bQ_+ = (0,0,0)$ and $\bQ_- = (\pi,\pi,0)$. 
The ratio of $I_{1}/I_{2}$ is chosen to give us the same value of the RKKY interaction in the cluster, 
$\bar{I}(\bQ)$ (defined as $\bar{I}(\bQ)=2/N\sum_{\tbq}I_{\tbq+\bQ}$). Again we fix $I_{\bQ_{AF}} = 2I$ for both cases. 

Here we provide the form of  $\rho(\bQ,\omega)$ we use at only $\bQ_{-}$; for the low-energy behavior, we will only need to
 keep the bosonic bath at $\bQ=\bQ_{-}$.
\begin{eqnarray}
\rho^{\rm type-I}_{I}(\bQ{-},\omega)&=& \frac{\Theta( 2I_{2}-\omega) \Theta(2I_{1}+2I_{2}+\omega)}{2I_{1}+4I_{2}} \\
\rho^{\rm type-II}_{I}(\bQ{-},\omega)&=&  \frac{\Theta( \omega- c+r)\Theta(c-\omega) \sqrt{r^{2}-(\omega-c)^{2}}}{\pi r^{2}/4}
\end{eqnarray}
where, for $\rho^{type-II}_{I}$ ,we have used the parameterization $c=2 I_{2}$, $r= -2 I_{1} + 8 I_{2}$. 
They represent the same shape as in equation (\ref{eq:rho2d}) (\ref{eq:rho3d}) for the Ising case but shifted along the $\epsilon$ axis.

Within the C-EDMFT approach, 
 the single particle and spin self energies 
 depend on cluster momentum $\bQ$, which yields a spin susceptibility 
\begin{equation}
\chi(\tbq+\bQ,i\omega_n,T) = \frac{1}{I_{\tbq + \bQ}+M(\bQ,i\omega_n)}.
\end{equation}
To avoid double-counting the RKKY interactions,
 the conduction electron bath does not become polarized by the finite magnetic order parameter~\cite{pixley2015cluster}. This  
 is achieved by taking a featureless density of states for the conduction band 
\begin{equation}
\rho_c(\epsilon) =\rho_0\Theta(D-|\epsilon|),
\end{equation}
for a half bandwidth $D$ and
self-consistently solve for the bosonic baths (see references~\cite{si2005magnetic,glossop2007magnetic,zhu2007zero} for the
 one site case).
Likewise, because the RKKY interaction is explicitly included
 (via $I_{ij}$), 
 we drop the dynamic inter-impurity interaction in the lattice model. 
  This is achieved within the effective cluster model, by taking the two impurities to be infinitely far apart~\cite{pixley2015cluster}, 
  and they are then only coupled by $\bar{I}_{{\bf Q}}$ and the bosonic baths. 
  This then corresponds to the cluster Hamiltonian 
\begin{eqnarray}
H_C &=& \sum_{i=1,2}H_{AI}^i 
+\sum_{\bQ,m}\bar{I}_{\bQ}S^{m}_{\bQ,f}S^{m}_{\bQ,f} + \sum_{\tbq, \bQ,m}\omega_{\tbq, \bQ}^{m}\phi^{m \dag}_{\tbq, \bQ}\phi^{m}_{\tbq, \bQ}
\nonumber
\\
&+& 
\sum_{\tbq, \bQ,m}g^{m}_{{\bf Q}}(\tbq)S^{m}_{{\bf Q},f}(\phi^{m \dag}_{\tbq, \bQ}+\phi^{m}_{-\tbq, -\bQ}) +\hloc S_{\bQ,f}^z.
\end{eqnarray}
where ${\bf Q}={\bf Q_{\pm}}$, $\tilde{\bf q}$ run though the momentum points in each patches, and $m=z$ for the ising anisotropic case and $m=x,y,z$ for the SU(2) symmetric case. In the last term, we have included $\hloc=[\delta I_{{\bf Q}_{AF}}+\chi^{z}_{0,-}(i\omega_n =0)]\langle S_{{\bf Q}_{AF},f}^z \rangle$, with $\delta I_{{\bf Q}_{AF}}= I_{{\bf Q}_{AF}}- \bar{I}_{{\bf Q}_{AF}}$, to incorporate antiferromagnetic order.  This then determines the magnetic order parameter $M_{AF} =  \langle S_{{\bf Q}_{AF},f}^z \rangle/\sqrt{2}$, where the spin operators in cluster momentum are $S^z_{\bQ_{\pm},f} = (S_{1,f}^z \pm S_{2,f}^z)/\sqrt{2}$. The Greens function of the bosonic baths give rise to the dynamic Weiss fields through
\begin{equation}
{\chi_{0,\bQ}^{m}}^{-1}(i\omega_n) = \sum_{\tbq} \frac{2g^{m}_{{\bf Q}}(\tbq)^2\omega^{m}_{\tbq,\bQ}}{ (\omega_{\tbq,\bQ}^{m})^{2} - (i\omega_n)^2},
\end{equation}
where $g^{m}_{{\bf Q}}$, and $\omega^{m}_{\bq,\bQ}$ are determined self consistently.
Due to the coarse graining, the relevant energy scale for the RKKY interaction is the inter-site interaction at the ordering wave vector, which we take to be $I_{{\bf Q}_{AF}} \equiv - 2 I$ (note, this serves as a definition of $I$).   Each cluster spin $S^{m}_{\bQ_{\pm},f}$ couples to two self consistent bosonic baths that represent ferro- $(\phi^{m}_{\tbq,\bQ_+})$ and antiferro- $(\phi^{m}_{\tbq,\bQ_-})$ magnetic fluctuations in the lattice model.
  We have defined $H_{AI}^i$ with $i=1,2$, which are two independent Anderson impurity models (as a result of taking the infinite separation limit that eliminates the sign problem~\cite{pixley2015pairing}),
\begin{eqnarray}
H^i_{AI} &=& \sum_{\kk, \sigma} \bigl[ \epsilon_{\kk} c_{\kk i\sigma}^{\dag} c_{\kk i\sigma}
+ V \bigl(  f_{i\sigma}^{\dag}c_{\kk i\sigma} + \text{H.c.}\bigr) \bigr]
+ \epsilon_f n_{i,f}  + U n_{i\uparrow,f}n_{i\downarrow,f},
\end{eqnarray}
 and $n_{i\sigma,f}=f_{i\sigma}^{\dag}f_{i\sigma}$, $n_{i,f}=\sum_{\sigma}n_{i\sigma,f}$.  We take the Anderson parameters of each impurity to be the same and therefore each has the same Kondo temperature.  For $I=0$ the two impurities are independent, and we can characterize the bare Kondo temperature
  through $T_K^{0}\equiv1/\chi_{\mathrm{loc},i}(T\rightarrow 0)$, where $\chi_{\mathrm{loc},i} = \int_0^{\hbar/k_BT}d\tau \langle T_{\tau} S^z_i(\tau)S^z_i \rangle$ is the local static spin susceptibility of impurity $i$. 
  We fix $U=0.25 D = -2\epsilon_f$ at particle hole symmetry, and take a hybridization $\Gamma_0 = \pi \rho_0 |V|^2=0.25D.$  This leads to a relatively high bare Kondo temperature $T_K^{0}\approx 1.0 D$, where a high $T_K^{0}$ is advantageous to try and reach the quantum critical regime (which is quite challenging~\cite{pixley2013quantum}).

Within the C-EDMFT formalism the local spin susceptibilities $\chi_{\loc}(\bQ,\tau) = \langle T_{\tau} S^z_{\bQ}(\tau)S^z_{-\bQ}\rangle$ are related to the lattice susceptibility through the self consistent equation
\begin{eqnarray}
\chi_{\loc}(\bQ,i\omega_n)&=&\frac{N_c}{N}\sum_{\tbq}\chi(\bQ+\tbq,i\omega_n)
\nonumber
\\
&=&\int d\epsilon \frac{ \rho_I(\bQ,\epsilon)}{\epsilon+M(\bQ,i\omega_n)}. 
\end{eqnarray}
For 
type-I model, we define the local Kondo energy scale in the lattice model as 
\begin{equation}
E_{\loc,{\rm type-I}}= \lim_{T\rightarrow 0}T_K^0\exp\left[- \frac{2I}{\alpha} \chi_{\loc}(\bQ_{-},T) \right],
\label{eqn:Eloc}
\end{equation}
as described in Ref.\,\cite{si2001locally,si2003local} and $\alpha$ is defined in Eq.\,(\ref{eqn:chiaf}). 
Whereas in
type-II model,
 the local Kondo energy scale is nonzero at the QCP.
 This leads to the following functional form~\cite{si2001locally,si2003local}, 
 which we use to fit the numerical data of $\chi_{\loc}(\bQ_{-},i\omega_n,T)$ to 
\begin{equation}
 a-\sqrt{8}/(\tilde{E}_{\loc}(T)\sqrt{I})\sqrt{\omega_n}-2/\tilde{E}_{\loc}(T)^2\omega_n+b \omega_n^{3/2}
\end{equation}
where $a,b$ and $\tilde{E}_{\loc}(T)$ are fitting parameters.
 Extrapolating to zero temperature yields $\tilde{E}_{\loc, {\rm type-II}}=\tilde{E}_{\loc}(T\rightarrow 0)$ to zero temperature. 
 However, this definition of the local energy scale has an overall arbitrary scale factor, 
 namely $E_{loc,3d}=\mathcal{A}\tilde{E}_{\loc,{\rm type-II}}$.  
 We fix the overall scale by determining the low energy scale where $\chi_{\loc}(\bQ_{-},i\omega_n,T)$ 
 fails to follow the leading $\omega_n^{1/2}$ behavior.   
 The functional form of the fit has been obtained analytically for single site EDMFT in Ref.\,\cite{si2001locally,si2003local} 
 in the long wavelength limit, and we have included the fit parameter $b$ to extend to higher energies.

\noindent {\bf Type-I model}\\
Solving the model for a
type-I RKKY density of states in the Ising limit we arrive at the finite temperature phase diagram shown in
 Fig.\,\ref{fig:phase_diagram}(a).  
 For a small ratio $I/T_K^0$ we find a heavy Fermi liquid (FL) phase with a finite temperature crossover at $E_{\loc}$ to the quantum critical 
 non-Fermi liquid (NFL) regime which then gives way to an AF phase 
 for a large $I/T_K^0$. The finite temperature magnetic phase boundary is given by the N\'eel temperature $T_N$ 
 where $M_{AF}$ develops a finite value. Extrapolating $T_N$ to zero temperature, 
 yields a critical value for the $T=0$ transition at $I_c/T_K^0 = 0.42(1)$.  
 As shown in Fig.\,\ref{fig:chi_SC_and_M}(a),
 at zero temperature $\chi_{AF}$ diverges when the order parameter becomes finite, 
 proving that the $T=0$ transition is second order. In Fig.\,\ref{fig:static-and-dyn-chi_AF}(a),
 we show the temperature dependence of lattice susceptibilities $\chi_{AF}$.  At the QCP we find  
a power-law temperature dependence
with a critical exponent $\alpha_{\mathrm{Ising}} = 0.81(4)$ in good agreement with a value of $\alpha \approx 0.75$ 
as found in experiments on CeCu$_{6-x}$Au$_x$ \cite{schroder2000onset}, 
as well as previous numerical result from EDMFT \cite{grempel2003locally, zhu2003continuous, zhu2007zero, glossop2007magnetic} 
where $\alpha$ is found to be from 0.72 to 
0.78 depending on specific implementations.
As a result of the self-consistent equation
we find in
the type-I model $\chi_{\loc}({\bf Q}-,i\omega_{n}=0,T)=1/(2I)\log|1+2I \chi_{AF}(T)|$, therefore the diverging lattice susceptibility
implies logarithmically diverging local spin susceptibility.
Correspondingly, using Eq.\,(\ref{eqn:Eloc}), a diverging $\chi_{\loc}$ implies that the local Kondo energy scale $E_{\loc}$ vanishes, 
which means that the Kondo effect is critically destroyed at the QCP.  
We therefore conclude the 
type-I solution within C-EDMFT yields a Kondo destruction QCP, with a critical exponent consistent 
with the value experimentally measured on CeCu$_{6-x}$Au$_x$.

\noindent {\bf Type-II model}\\
Solving the model with a 
type-II RKKY density of states yields 
a nonzero renormalized Kondo energy scale at the QCP (see Fig.\,\ref{fig:Ising-sdw}).
Similar to the
type-I case we find an 
AF phase boundary $T_N$ where $M_{AF}$ becomes finite and $\chi_{AF}$ diverges.  Extrapolating $T_N$ to zero temperature yields a QCP at $I_c/T_K^0=0.42(1)$.  However, in contrast to the
type-I case the finite value of $E_{\loc}$ at the QCP can be seen directly in the self consistent equation for $\chi_{\loc}({\bf Q}_{-},i\omega_n,T)$, where the static susceptibility takes a finite value at the QCP and $E_{\loc}/T_K^0$ remains non-zero. However, due to the dynamical RKKY-Kondo competition,
it is  expected---and indeed found---to be small.

\noindent {\bf Static lattice pairing susceptibility}\\
We now turn to studying the strength of the superconducting pairing correlations between the correlated $4f$ electrons in the vicinity of a QCP.  
We do this by calculating the static lattice pairing susceptibility defined as
\begin{eqnarray}
\chi_{SC}(T)=\frac{1}{N(z/2)}\sum_{\langle i,j\rangle,\sigma}\sum_{\langle k,l\rangle,\lambda}f^*_{i,j}f_{k,l}g^*_{\sigma\bar{\sigma}}g_{\lambda\bar{\lambda}}
\int_0^{\hbar/k_BT}d\tau \langle T_{\tau} \hat{\Delta}_{i\sigma j\bar{\sigma}}(\tau)\hat{\Delta}^{\dag}_{k\lambda l\bar{\lambda}}\rangle,
\label{eq:chisc_real}
\end{eqnarray}
where we have introduced the pair operator $\hat{\Delta}_{i\sigma j\bar{\sigma}}=f_{i\sigma}f_{j\bar{\sigma}}$. In addition,
 $N(z/2)$ is the number of bonds in the lattice (and serves as a normalization), 
 with $N$ being the total number of sites and $z$ being the number of nearest neighbors. 
 We project the pairing susceptibility into different symmetry channels through 
the symmetry factor in real space $f_{ij}$ and that in spin space $g_{\sigma\bar{\sigma}}$ (see ref.~\cite{Mineev-book}).  
In the following we are only concerned with the spin-singlet superconductivity,
which is given by $g_{\uparrow\downarrow/\downarrow\uparrow}=\pm 1$. 
Restricting to a two site cluster, we 
consider  $f_{ij}=1$ for $i$ and $j$ being nearest neighbors and zero otherwise. 
The two site cluster EDMFT only distinguishes spin singlet vs triplet pairing symmetries; 
extended s-wave and d-wave susceptibilities are indistinguishable within the current 2-site cluster approximation; a minimum of four site cluster is needed to resolve this, which is left for future work.
Within our approach $\chi_{SC}$ is obtained by solving for the irreducible vertex function in the particle-particle channel of the Bethe-Salpeter equation 
in the cluster. In turn, using the vertex function and the bare particle-particle bubble,
we can construct the lattice pairing correlation function. 
We consider the case of the RKKY interaction being 
SU(2) symmetric. 

\noindent {\bf Calculation of the pairing susceptibility}\\
We now discuss the calculation of the lattice pairing susceptibility. This is most conveniently formulated in the momentum space.
The dynamical lattice pairing susceptibility is defined as \cite{XiChen-PRL},
\begin{eqnarray}
\chi_{\ud} ^{\nu_{n}\nu_{n}^{\prime},kk^{\prime}}(\omega_{n}, q) 
&=&\int d \tau_{1} d\tau_{2}d\tau_{3}d\tau_{4} e^{i \nu_{n} (\tau_{3}-\tau_{1})}  e^{i \nu_{n}^{\prime} (\tau_{4}-\tau_{2})} e^{i \omega_{n}(\tau_{2}-\tau_{3}) } \nonumber \\
&\times& \langle  T_{\tau} c_{k+q\uparrow}^{\dagger} (\tau_{1})  c_{-k^{\prime} \downarrow} (\tau_{2}) c_{-k \downarrow}^{\dagger} (\tau_{3} ) c_{k^{\prime}+q \uparrow}(\tau_{4}) \rangle 
\label{dynamical_pair_formula}
\end{eqnarray}
Here $\nu_{n}$ and $\nu^{\prime}_{n}$ are fermionic matsubara frequencies, and $\omega_{n}$ are bosonic matsubara frequencies.
It is related to the static lattice pairing susceptibility 
defined earlier in Eq.\,(\ref{eq:chisc_real}) through \cite{Jarrell-PRB},
\begin{equation}
\chi_{SC}(T)=-\frac{1}{\beta^{2} } \sum_{\nu_{n} \nu^{\prime}_{n}, k k^{\prime}} \chi^{ \nu_{n} \nu^{\prime}_{n},k k^{\prime} } _{\ud} (\omega_{n}=0, q=0 )  f(k)f(k^{\prime})
\end{equation}
with $f(k)=\cos(k_{x})+\cos(k_{y})$ (type-I) or $f(k)=\cos(k_{x})+\cos(k_{y})+\cos(k_{z})$ (type-II) being the pairing form factor in the momentum space. 
Since we only focus on the $\omega_{n}=0, q=0$ case in the remaining part we will drop these two indices.

The dynamical pairing susceptibility is given by \cite{Rohringer-PRB},
\begin{eqnarray}
\chi_{\ud}^{\nu_{n} \nu_{n}^{\prime},k k^{\prime} }  
=\chi_{0,\ud}^{\nu_{n} \nu_{n}^{\prime}, k k^{\prime} } 
-\frac{1}{\beta^{2}} \sum_{ \substack{\nu_{1n}\nu_{2n} \\ k_{1} k_{2} } }  \chi_{0,\ud}^{\nu_{n} \nu_{1n},k k_{1} } 
F^{\nu_{1n}\nu_{2n},k_{1}k_{2} }_{\udb}  \chi_{0,\ud}^{\nu_{2n} \nu_{n}^{\prime}, k_{2} k^{\prime} }
\label{eq:chi_F}
\end{eqnarray}
where $F^{\nu_{1n}\nu_{2n},k_{1}k_{2} }_{\udb}  $ is the full vertex function in the particle-particle channel,  and $\chi_{\ud,0}^{\nu_{1n} \nu_{2n} , k_{1} k_{2} } $ is the bare particle-particle bubble given by the single particle Greens function 
$\chi_{0,\ud}^{\nu_{1n} \nu_{2n} , k_{1} k_{2} } = -\beta G_{\uparrow}(k_{1}, \nu_{1n} ) G_{\downarrow}(-k_{2} ,-\nu_{2n}) \delta_{\nu_{1n},\nu_{2n}}\delta_{k_{1},k_{2}}$. 
We will also use the shorthand notation that $\udb$ stands for $\uparrow\downarrow\downarrow\uparrow$ and $\ud$ stands for $\uparrow\uparrow\downarrow\downarrow$.

From the Bethe Salpeter equation, the full vertex can be expressed in terms of the irreducible vertex \cite{Rohringer-PRB}.
\begin{eqnarray}
F_{\udb}^{\nu_{n} \nu_{n}^{\prime}, k k^{\prime} } 
=
\Gamma_{\udb}^{\nu_{n} \nu_{n}^{\prime} ,k k^{\prime} } 
&+&
 \frac{1}{2\beta^{2}}  \sum_{ \substack{\nu_{1n}\nu_{2n} \\ k_{1} k_{2} } } 
\Gamma_{\udb}^{\nu_{2n} \nu_{n}^{\prime} ,k_{2} k^{\prime} }
\chi_{0,\ud}^{\nu_{1n} \nu_{2n},k_{1} k_{2} }  F_{\ud}^{\nu_{n} (-\nu_{1n}), k (-k_{1}) } \nonumber \\
&+&
 \frac{1}{2\beta^{2}}   \sum_{ \substack{\nu_{1n}\nu_{2n} \\ k_{1} k_{2} } } 
\Gamma_{\dub}^{\nu_{2n} \nu_{n}^{\prime} , k_{2} k^{\prime} }
\chi_{0,\du}^{(-\nu_{1n})( -\nu_{2n}), (-k_{1}) (-k_{2}) }  F_{\du}^{\nu_{n} (-\nu_{1n}), k (-k_{1}) } 
\end{eqnarray}

When the model has SU(2) symmetry, we can utilize the relation 
$\Gamma_{\du}^{\nu_{n} \nu_{n}^{\prime}, k k^{\prime}  } 
  {=} \Gamma_{\uparrow\downarrow}^{\nu_{n} \nu_{n}^{\prime}, k k^{\prime} }  $,  
$\chi_{0,\ud}^{\nu_{n} \nu_{n}^{\prime},k k^{\prime} } =\chi_{0,\du}^{\nu_{n} \nu_{n}^{\prime},k k^{\prime} } $ and use crossing symmetry \cite{Rohringer-PRB}
$F_{\ud} ^{\nu_{n} (-\nu^{\prime}_{n}), k  (-k^{\prime}) } 
 {=}
 - F_{\overline{\uparrow\downarrow}}^{\nu_{n}\nu_{n}^{\prime} , k k^{\prime} }$
to simplify the above equation,
\begin{eqnarray}
F_{\overline{\uparrow\downarrow}}^{\nu_{n} \nu_{n}^{\prime}, k k^{\prime} }
&=&
\Gamma_{\overline{\uparrow\downarrow}}^{\nu_{n}\nu_{n}^{\prime}, k k^{\prime} } 
-\frac{1}{\beta^{2} }  \sum_{ \substack{\nu_{1n}\nu_{2n} \\ k_{1} k_{2} } } 
\Gamma_{\overline{\uparrow\downarrow}}^{\nu_{2n} \nu_{n}^{\prime}, k_{2} k^{\prime} } 
\chi_{0,\ud}^{\nu_{1n} \nu_{2n}, k_{1} k_{2}  }  F_{\overline{\uparrow\downarrow}}^{\nu_{n} \nu_{1n}, k k_{1} }
\label{eq:F_Gamma}
\end{eqnarray}
From Eq.\,(\ref{eq:chi_F}) and Eq.\,(\ref{eq:F_Gamma}) we can eliminate 
$ F_{\overline{\uparrow\downarrow}}^{\nu_{n} \nu_{n}^{\prime}, k k^{\prime} }$ and obtain
\begin{equation}
\chi_{\ud}^{\nu_{n} \nu_{n}^{\prime} , k k^{\prime} } = \chi_{0,\ud}^{\nu_{n} \nu_{n}^{\prime} , k k^{\prime} }
-\frac{1}{\beta^{2}} \sum_{ \substack{\nu_{1n}\nu_{2n} \\ k_{1} k_{2} } }  \chi_{0,\ud}^{\nu_{n} \nu_{1n} , k k_{1} } \Gamma^{\nu_{n1}\nu_{2n},k_{1}k_{2}}_{\udb}  \chi_{0,\ud}^{\nu_{2n} \nu_{n}^{\prime},k_{2} k^{\prime} }
\label{eq:chipairlattice}
\end{equation}

Withinquantum cluster theories \cite{Jarrell-PRB} \cite{Maier-PRL}, the irreducible vertex is 
depends on cluster momentum,
\begin{equation}
 \Gamma^{\nu_{n} \nu_{n}^{\prime},kk^{\prime}}_{\udb}= \Gamma^{\nu_{n}\nu_{n}^{\prime},KK^{\prime}}_{\udb}
\end{equation}

Defining $\chi_{c,\ud}^{\nu \nu^{\prime} , k k^{\prime} }$ and $\chi_{c0,\ud}^{\nu \nu^{\prime} , k k^{\prime} }$
 as the corresponding cluster quantities for $\chi_{\ud}^{\nu \nu^{\prime} , k k^{\prime} }$ and $\chi_{0,\ud}^{\nu \nu^{\prime} , k k^{\prime} }$, 
 we have the analogous equation for the cluster quantities (notice that they share the same irreducible vertex due to the approximation we have made).
\begin{eqnarray}
\chi_{c,\ud}^{\nu_{n} \nu_{n}^{\prime} , K K^{\prime} } &=& \chi_{c0,\ud}^{\nu_{n} \nu_{n}^{\prime} , K K^{\prime} }
-\frac{1}{\beta^{2}} \sum_{ \substack{\nu_{1n}\nu_{2n} \\ k_{1} k_{2} } }  \chi_{c,\ud}^{\nu_{n}\nu_{1n},K K_{1} }  \Gamma^{\nu_{1n}\nu_{2n},K_{1}K_{2}}_{\udb}  \chi_{c0,\ud}^{\nu_{2n} \nu^{\prime},K_{2} K^{\prime} }
\label{eq:chipaircluster}
\end{eqnarray}

Our strategy is to obtain $\chi_{c,\ud}^{\nu_{n} \nu_{n} ^{\prime} , K K^{\prime} }$ and $\chi_{c0,\ud}^{\nu_{n} \nu_{n}^{\prime} , K K^{\prime} }$ from the cluster model in the converged C-EDMFT solution, from which we obtain $\Gamma^{\nu_{n}\nu_{n}^{\prime},KK^{\prime}}_{\udb}$ using Eq.\,(\ref{eq:chipaircluster}) 
and finally get $\chi_{\ud}^{\nu_{n} \nu_{n}^{\prime},k k^{\prime} }$ using Eq.\,(\ref{eq:chipairlattice}). In CT-QMC,
 the calculation of  $\chi_{c,\ud}^{\nu_{n} \nu_{n} ^{\prime} , K K^{\prime} }$ is 
 achieved using worm algorithm \cite{Gunacker-2015,Gunacker-2016}.

To obtain the single particle Greens function for the bare particle bubble, we use the Dyson equation,
\begin{equation}
{G}^{-1}(k,i\nu_{n})={G}_{0}^{-1}(k,i\nu_{n})-{\Sigma}(k,i\nu_{n})
\end{equation}
and coarse-grain the single particle self-energy
\begin{equation}
{\Sigma}(k,i\nu_{n})={\Sigma}(K,i\nu_{n})
\end{equation}
which is obtained from the cluster Dyson equation
\begin{equation}
{G}_{c}^{-1}(K,i\nu_{n})={G}_{0,c}^{-1}(K,i\nu_{n})-{\Sigma}(K,i\nu_{n})
\end{equation}

The lattice non-interacting Greens function for the lattice and the cluster is given by,
\begin{eqnarray}
\hat{G}_{0}^{-1}(k,i\nu_{n})&=&\frac{1}{i\nu_{n} + \mu - \frac{ 2 \Gamma_{0}^{lat}/\pi }{i\nu_{n} - \epsilon_{k}}} \\
\hat{G}_{0,c}^{-1}(K,i\nu_{n})&=&\frac{1}{i\nu_{n} + \mu - \int_{-D}^{D} d \epsilon \frac{ 2 \Gamma_{0}/\pi }{i\nu_{n} - \epsilon}}
\end{eqnarray}

Following the prescription for the fermionic section in C-EDMFT, 
we do not enforce the following self-consistency equation for the single particle Greens function,
\begin{equation}
G_{c}(K,i\nu_{n})=\sum_{\tilde{k}} G(K+\tilde{k},i\nu_{n} )
\end{equation}
Instead we choose an
effective hybridization parameter $\Gamma_{0}^{lat}=0.6\Gamma_{0}$ for the lattice non-interacting 
Greens function so that the above relation will hold 
to a good approximation.

The component in the bare particle $\hat{\chi}_{0,\ud}^{\nu_{n}\nu_{n}^{\prime},k k^{\prime}}$ containing the anomalous Greens function is given by,
 \begin{eqnarray}
 \chi^{\nu_{n}\nu_{n}^{\prime},k k^{\prime}+Q_{AF}}_{0,\ud} &=& G_{\uparrow}(k+Q_{AF},k,i\nu_{n}) G_{\downarrow}(-k-Q_{AF},-k,-i\nu_{n}) \nonumber \\
  \chi^{\nu_{n}\nu_{n}^{\prime},k+Q_{AF} k^{\prime}}_{0,\ud} &=& G_{\uparrow}(k,k+Q_{AF},i\nu_{n}) G_{\downarrow}(-k,-k-Q_{AF},-i\nu_{n}) 
 \end{eqnarray}
 where $G_{\sigma}(k+Q_{AF},k,i\nu_{n})=\int_{0}^{\beta} d\tau \langle T_{\tau} c_{K,\sigma}(\tau) c_{K+Q_{AF},\sigma}^{\dagger} \rangle$.
 
 After manipulation using the crossing relation and $\chi_{0, \downarrow \uparrow }^{(-\nu_{n}) (-\nu_{n}^{\prime}),(-k)( - k^{\prime})}
=\chi_{0, \uparrow \downarrow }^{\nu_{n} \nu_{n}^{\prime},k k^{\prime}} $ (which only hold in the special case of $\omega=0,q=0$), we can write the
Bethe-Salpeter equation in a compact matrix form,
\begin{eqnarray}
\hat{F}_{\udb}^{\nu_{n} \nu_{n}^{\prime} , k k^{\prime} } 
=\hat{\Gamma}_{\udb}^{\nu_{n} \nu_{n}^{\prime} , k k^{\prime} }\nonumber 
&-&\frac{1}{2 \beta^{2}}\sum_{\nu_{1n},\nu_{2n} }  \hat{F}_{\udb}^{\nu_{n} \nu_{1n}, k k_{1} } \hat{\chi}_{0,\ud}^{\nu_{1n} \nu_{2n}, k_{1} k_{2}  }  \hat{\Gamma}_{\udb}^{\nu_{2n} \nu_{n}^{\prime}, k_{2} k^{\prime} } \nonumber  \\
&-&\frac{1}{2 \beta^{2}}\sum_{\nu_{1n},\nu_{2n} }  \hat{F}_{\udb}^{\nu_{n} \nu_{1n}, k k_{1} } \hat{\chi}_{0,\ud}^{\nu_{1n} \nu_{2n}, k_{1} k_{2}  }  \hat{\Gamma}_{\dub}^{\nu_{2n} \nu_{n}^{\prime}, k_{2} k^{\prime} }  
\end{eqnarray}
\begin{equation}
\hat{\chi}_{\ud}^{\nu_{n}\nu_{n}^{\prime},k k ^{\prime}} =\hat{\chi}^{\nu_{n}\nu_{n}^{\prime},k k ^{\prime}}_{0 ,\ud}
- \hat{\chi}^{\nu_{n}\nu_{1n},k k_{1} }_{0 ,\ud} 
\hat{F}_{\ud}^{\nu_{1n} \nu_{2n},k_{1} k_{2} }
\hat{\chi}_{0 ,\ud}^{\nu_{2n} \nu_{n}^{\prime},k_{2} k ^{\prime}}
\end{equation}

where
\begin{equation}
\hat{F}_{\udb}^{\nu_{n}\nu_{n}^{\prime},k k^{\prime}}
=\begin{pmatrix}
F^{\nu_{n}\nu_{n}^{\prime},k k^{\prime}}_{\udb} & F^{\nu_{n}\nu_{n}^{\prime},k k^{\prime}+Q_{AF}}_{\udb} \\
F^{\nu_{n}\nu_{n}^{\prime},k+Q_{AF} k^{\prime}}_{\udb} & F^{\nu_{n}\nu_{n}^{\prime},k+Q_{AF} k^{\prime}+Q_{AF}}_{\udb} 
\end{pmatrix}
\end{equation}
and similarly for $\hat{\Gamma}_{\udb}^{\nu_{n}\nu_{n}^{\prime},k k^{\prime}}$,and $\hat{\chi}_{\ud}^{\nu_{n}\nu_{n}^{\prime},k k^{\prime}}$.

\noindent
{\bf Numerical implementation}\\
 For the Ising case, using the fact that the two bosonic baths commute, we can apply the CT-QMC approach used in Ref.\,\onlinecite{pixley2015pairing} to solve the cluster model.  
 For the SU(2) case, we use the CT-QMC approach in Ref.\,\onlinecite{cai2019bose,otsuki2013spin,steiner2015double} to deal with the three component 
 vector bosonic bath and the Heisenberg spin-spin interaction in the cluster model. 
 For the low energy physics, we focus on the coupling to the antiferromagnetic bosonic bath. This is fortuitous, 
 as the coupling to the ferromagnetic bosonic bath 
 will introduce a sign problem.


\clearpage
\begin{figure}[h!]
\center
\includegraphics[height=3.8in,angle=0]{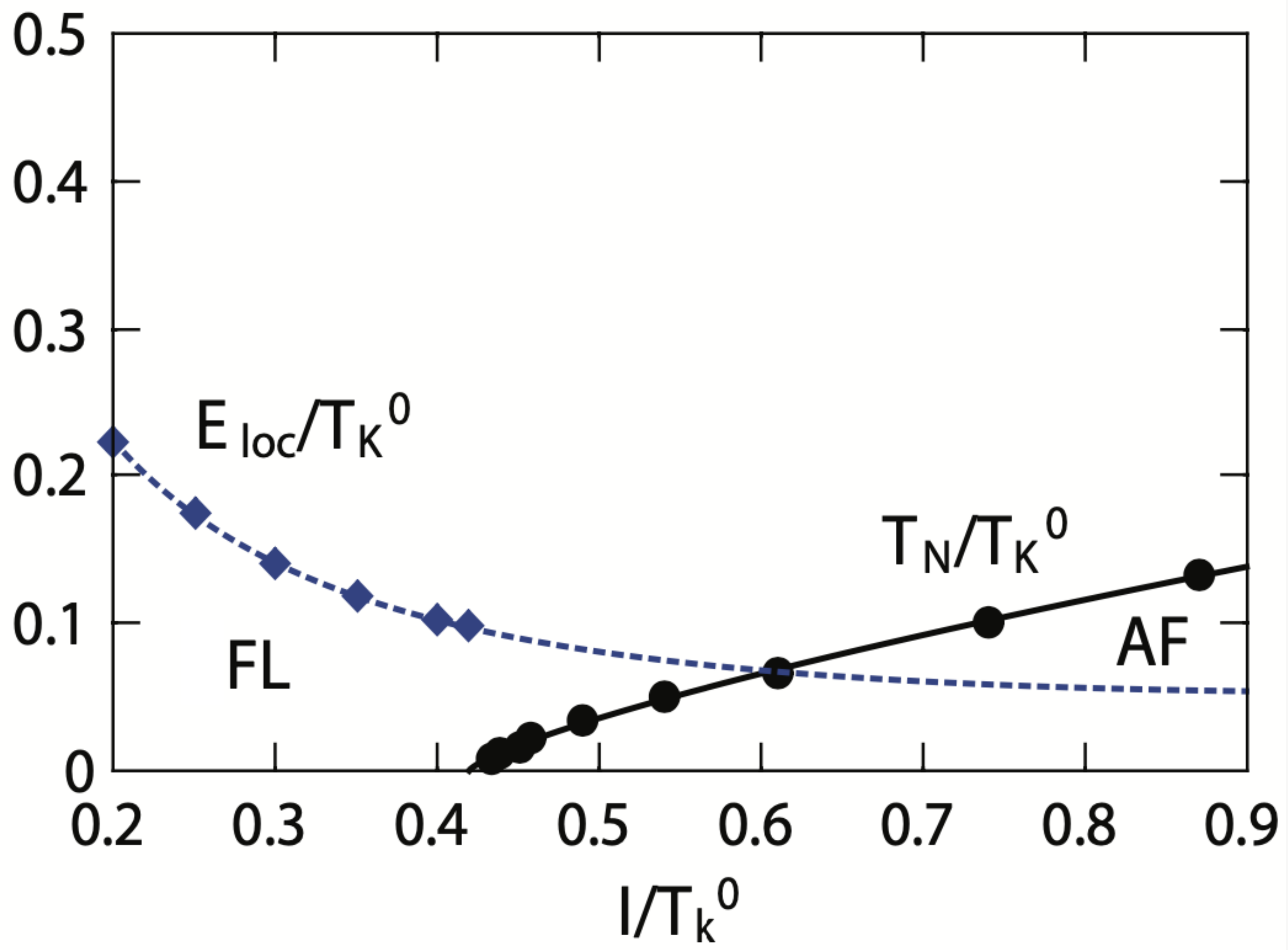}
\caption{Finite temperature phase diagram vs the ratio of $I/T_K^0$ for the type-II model with Ising anisotropy.  
For small $I/T_K^0$ we find the model is in the FL phase characterized by the static spin susceptibility saturating 
to a constant for $T \ll E_{\loc}$.  
For large $I/T_K^0$ the model develops antiferromagnetic (AF) order.  
In between the two is the quantum critical non-Fermi liquid regime, where $E_{cr} \equiv E_{\loc}(\delta_c) $ remains nonzero but 
is small
compared to $T_K^0$.}
\label{fig:Ising-sdw}
\end{figure}

\clearpage
\begin{figure}[h!]
\center
\includegraphics[height=2in,angle=0]{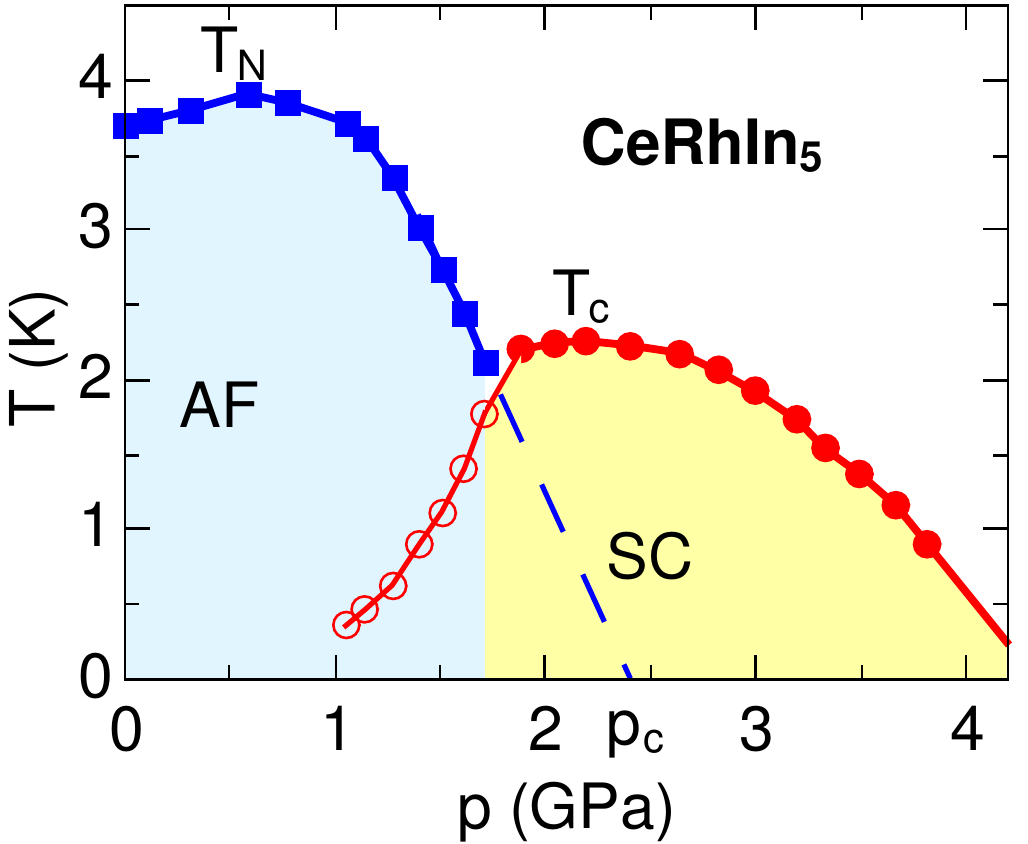}
\caption{Pressure-temperature phase diagram of CeRhIn$_{5}$ at zero magnetic field \cite{park2006hidden,Kne08.1}. 
$T_{N}$ is determined from specific heat measurements. $T_{c}$ is determined from specific heat measurements up to $p=2.5$Gpa and from resistivity measurement at higher pressure.}
\label{fig:exp}
\end{figure}

\end{document}